\documentclass[journal=jctcce,manuscript=article]{achemso}
\setkeys{acs}{usetitle=true}

\usepackage[version=3]{mhchem} 

\usepackage{amsmath,amssymb}
\usepackage{graphicx,color,subfigure}
\usepackage{amsfonts}
\usepackage{comment}
\usepackage{placeins}
\usepackage{multirow}

\usepackage{sectsty}
\usepackage{balance}
\usepackage{xspace}
\usepackage{algorithm,algorithmic}

\usepackage{appendix}
\usepackage{float}

\usepackage{epsfig,amssymb,amsmath,amsthm,amsfonts,amsbsy,mathrsfs}
\usepackage{graphicx,color}
\usepackage{amsmath}
\usepackage{amssymb}
\usepackage{epstopdf}

\usepackage{graphicx} 
\graphicspath{{figure/}}
\usepackage{lastpage}
\usepackage[format=plain,justification=raggedright,singlelinecheck=false,font=small,labelfont=bf,labelsep=space]{caption}

\allowdisplaybreaks

\usepackage[version=3]{mhchem} 

\newcommand{\vt}{\vec{\theta}}

\definecolor{MatlabY}{rgb}{0.929,0.694,0.125}

\author{Yanxian Tao}
\affiliation{Hefei National Research Center for Physical Sciences at the Microscale, University of Science and Technology of China, Hefei 230026, China}

\author{Xiongzhi Zeng}
\affiliation{Hefei National Research Center for Physical Sciences at the Microscale, University of Science and Technology of China, Hefei 230026, China}

\author{Yi Fan}
\affiliation{Hefei National Research Center for Physical Sciences at the Microscale, University of Science and Technology of China, Hefei 230026, China}

\author{Jie Liu}
\email{liujie86@ustc.edu.cn (Jie Liu)}
\affiliation{Hefei National Laboratory, University of Science and Technology of China, Hefei 230088, China}

\author{Zhenyu Li}
\affiliation{Hefei National Research Center for Physical Sciences at the Microscale, University of Science and Technology of China, Hefei 230026, China}
\alsoaffiliation{Hefei National Laboratory, University of Science and Technology of China, Hefei 230088, China}

\author{Jinlong Yang}
\email{jlyang@ustc.edu.cn (Jinlong Yang)}
\affiliation{Hefei National Research Center for Physical Sciences at the Microscale, University of Science and Technology of China, Hefei 230026, China}
\alsoaffiliation{Hefei National Laboratory, University of Science and Technology of China, Hefei 230088, China}

\title{Exploring accurate potential energy surfaces via integrating variational quantum eigensovler with machine learning}

\begin{document}

\begin{abstract}
The potential energy surface (PES) is crucial for interpreting a variety of chemical reaction processes. However, predicting accurate PESs with high-level electronic structure methods is a challenging task due to the high computational cost. As an appealing application of quantum computing, we show in this work that variational quantum algorithms can be integrated with machine learning (ML) techniques as a promising scheme for exploring accurate PESs. Different from using a ML model to represent the potential energy, we encode the molecular geometry information into a deep neural network (DNN) for representing parameters of the variational quantum eigensolver (VQE), leaving the PES to the wave function ansatz. Once the DNN model is trained, the variational optimization procedure that hinders the application of the VQE to complex systems is avoided and thus the evaluation of PESs is significantly accelerated. Numerical results demonstrate that a simple DNN model is able to reproduce accurate PESs for small molecules. 
\end{abstract}

The potential energy surface (PES) is of particular importance for the analysis of molecular structures and chemical reaction dynamics. The evaluation of PESs first requires a high-level electronic structure method for accurate energy prediction. Many recent efforts have been devoted to the development of state-of-the-art computational methods, such as multireference wave function methods~\cite{LisNacAqu18,Mat21}, density matrix renormalization group~\cite{ChaSha11}, selected configuration interaction~\cite{LiuHof16,TubFreLev20} and Quantum Monte Carlo,~\cite{DubMitJur16,TubLeeTak16} towards systematic improvement of computational accuracy. Recent advance in quantum computing provides a new route for exploring electronic structure of molecules and materials.~\cite{abrams1997simulation, peruzzo2014variational, o2016scalable, kandala2017hardware, shen2017quantum, bian2019quantum, DuXuPen10, ColRamDah18, AruAryBab20, HugGorRub22} Utilizing the quantum nature of quantum computer, such as superposition and entanglement, many quantum algorithms have been proposed to accurately solve electronic structure problems in polynomial time~\cite{Pre18,CaoRomOls19,McAEndAsp20,BauBraMot20,HeaFliCic21}. These quantum algorithms provide a reliable scheme to describe both strong and weak correlation effects, which is essential for constructing accurate PESs.

Currently, the variational quantum eigensolver (VQE), a hybrid quantum-classical algorithm, is one of leading techniques to solve electronic structure problems on near-term quantum devices.~\cite{WhiBiaAsp11, McCRomBab16, McCKimCar17, RomBabMcC18, ColRamDah18} The VQE computes a target function on a quantum computer and minimize this function on a classical computer. As such, it maintains the circuit depth short to mitigate errors at the expense of many measurements.~\cite{peruzzo2014variational, MalBabKiv16} The VQE adapts to available qubit counts, coherence time, and gate fidelity for a feasible evaluation of the target function on contemporary quantum hardware, while the nonlinear optimization of circuit parameters towards a global minimum of the target function remarkably complicates this algorithm. This hinders the application of the VQE to efficiently construct accurate PESs of complex systems. 

Due to powerful capabilities on data processing~\cite{hinton2006reducing, lecun2015deep}, machine learning (ML) techniques have been widely used for obtaining a numerically efficient and accurate representation of PESs by fitting and interpolating of large quantities of ab initio data.~\cite{behler2007generalized,jiang2016potential,chmiela2017machine,zhang2018deep,HuXieLi18,ChenLiuFan18,ZheZubWu21,ZheYanWu22}. A simple strategy for combining the VQE with a ML method, such as deep neural network (DNN), is to produce a set of training data from the VQE and train the ML model for predicting PESs. Here, the training data set is often composed of the energies and/or forces. For strongly-correlated and excited-state systems, the DNN model is expected to have sufficiently expressive power to represent accurate PESs.~\cite{WesMar21} In addition, the training set consisting of a huge number of geometries is necessary to describe (near-)degenerate cases.~\cite{chen2018deep,DraBarThi18} These may present a grand challenge for applying the ML methods to electronic structure problems. Thus, it is necessary to explore other strategies that can utilize the power of both quantum computers and ML techniques. 

In this work, we employ a DNN model to represent circuit parameters of the VQE, leaving the highly entangled electronic states to a parameterized wave function ansatz. Here, the DNN maps circuit parameters to a feature space, where the molecular geometry information are encoded in linear transformations with learnable parameters. Unitary coupled cluster with generalized single and double excitations (UCCGSD), which has been demonstrated to be able to produce exact PESs for small molecules,~\cite{LeeHugHea19,LiuWanLi20} is used to prepare electronic states. We apply this DNN-VQE method to several chemical molecules to explore their accurate PESs.

\begin{figure}[!htb]
\begin{center}
\includegraphics[width=0.8\textwidth]{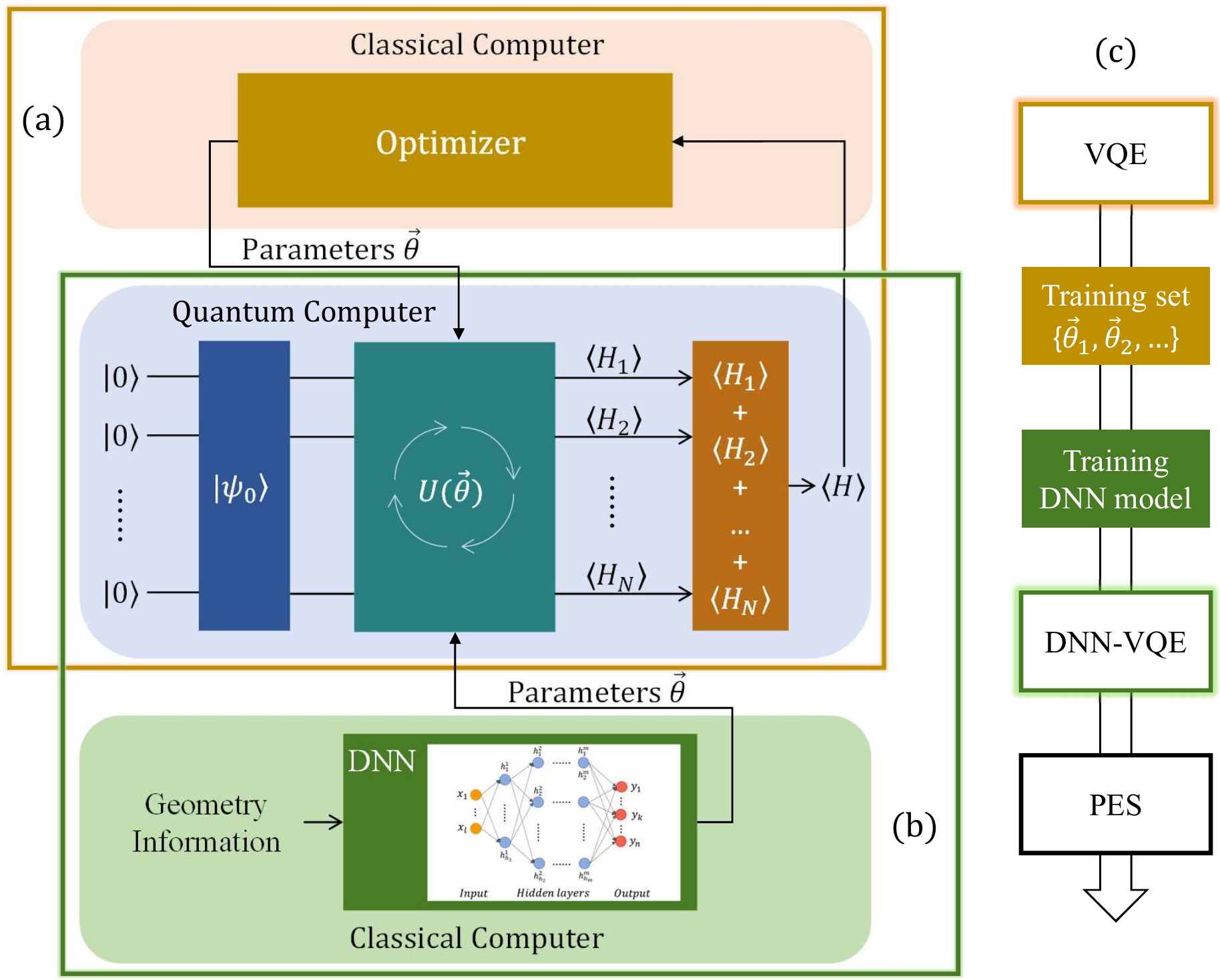}
\end{center}
\caption{(a) Schematic diagram of the standard VQE algorithm. The parameterized quantum circuit is trained with a classical optimizer; (b) Schematic diagram of the DNN-VQE method. The circuit parameters are represented by a deep neural network model; (c) Flowchart of constructing the potential energy surface.}
\label{fig:fig1}
\end{figure}

The VQE determines the eigenstates of the electronic Hamiltonian based on the Rayleigh-Ritz variational principle~\cite{peruzzo2014variational}. The wave function ansatz is written as 
\begin{equation}
    |\Psi(\vt)\rangle = \hat{U}(\vt) |\Psi_0\rangle,
\end{equation}
where $\hat{U}$ is a unitary transformation and $|\Psi_0\rangle$ is the reference state, which is often chosen to be Hartree-Fock (HF) wave function. The total energy is determined by
\begin{equation}
    E = \min_{\vt}  \langle \Psi(\vt) | \hat{H} | \Psi(\vt) \rangle.
\end{equation}
The Hamiltonian $\hat{H}$ in the second-quantized form is
\begin{equation}
        \hat{H} = \sum_{pq} v^p_q a^\dag_p a_q + \frac{1}{2} \sum_{pqrs} v^{pq}_{sr}  a^\dag_p a^\dag_q a_r a_s,
\end{equation}
where $v^p_{q}$ are the one-electron integrals and $v^{pq}_{rs}$ are the two-electron integrals. The overall procedure of the VQE is shown in Fig.~\ref{fig:fig1}(a). After state preparation, the total energy is measured on quantum devices and fed back to a classical optimizer to optimize circuits parameters. This procedure continues until the convergence is achieved.

The accuracy of the VQE depends heavily on the wave function ansatz. Unitary coupled cluster (UCC) is a widely used chemically inspired ansatz in the VQE algorithm~\cite{Kut82,BarKucNog89,TauBar06,Eva11,HarShiScu18}. Different from the traditional coupled cluster theory, the UCC defines a unitary transformation operator,
\begin{equation}\label{eq3}
    U(\vec{t}) = e^{\hat{T} - \hat{T}^\dag}.
\end{equation}
The cluster operator of UCCGSD is defined as
\begin{equation}\label{eq5}
    \hat{T} = \frac{1}{2}\sum_{pq} t^p_q a^\dag_p a_q + \frac{1}{4}\sum_{pqrs} t^{pq}_{rs} a^\dag_p a^\dag_q a_r a_s,
\end{equation}
where $p,q,\ldots$ indicate general orbitals. The coefficients $t^p_q$ and $t^{pq}_{rs}$ are to be variationally determined. Numerical studies of small molecules with minimum basis sets demonstrated that UCCGSD is far more robust and accurate than UCC with singles and doubles~\cite{LeeHugHea19}. In this work, we employ UCCGSD as the wave function ansatz to accurately describe the PESs of small molecules.

One notorious problem of the VQE is the highly nonlinear optimization of circuit parameters to find a global energy minimum, which is an open problem in mathematics. This problem has been discussed for the UCC and UpCCGSD ansatzes, in which many optimizations starting from random initial parameters are carried out to approximate the global minimum.~\cite{LeeHugHea19} For complex systems, a more complicated sampling scheme is probably necessary to approach the global minimum. It is evident that this optimization procedure could result in a significant increase of the computational cost in the evaluation of PESs. A simple strategy to alleviate this optimization problem is to introduce the ML techniques, such as DNN, to fit and interpolate circuit parameters.  

The DNN has attracted more and more attention in recent years for computing high-dimensional PESs.~\cite{WesMar21} A schematic model of the DNN~\cite{GooBenCou16} with multiple inputs and multiple outputs is shown in Fig.~\ref{fig:fig1}(b), where yellow, blue and red circles represent neurons of the input, hidden and output layers, respectively. And $h^a_b$ indicates the $b^{th}$ neuron in the $a^{th}$ hidden layer. The accuracy of the DNN depends on the input vectors, which are molecular descriptors that satisfy translational, rotational and permutational symmetry. In this work, the inputs $\vec{x} = (x_1, x_2, \dotsc, x_l)$ are the local geometry information, and the outputs $\vec{y} = (y_1, y_2, \dotsc, y_n)$ are non-zero variational parameters in the VQE. The output of the $k^{th}$ neuron in the first hidden layer is 
\begin{equation}
    h^1_k = f(\sum^l_{i=1}u_{ik} x_i + a^1_k),
\end{equation}
where $u_{ik}$ is the weight of $x_i \rightarrow h^1_k$, $a^1_k$ is the bias of $h^1_k$ neuron, $f$ indicates the ReLU activation function. In the following, the output of the $k^{th}$ neuron in the $m^{th}$ hidden layer is 
\begin{equation}
        h^m_k = f(\sum^{N_{m-1}}_{i=1} v^m_{ik} h^{m-1}_i + a^m_k),
\end{equation}
where $v^j_{ik}$ is the weight of $h^{j-1}_i \rightarrow h^j_k$, $a^j_k$ is the bias of $h^j_k$ neuron, and $f$ indicates the ReLU activation function. And the output of the DNN is
\begin{equation}
        y_k = \sum^{N_{m}}_{i=1} w_{ik} h^m_i + b_k,
\end{equation}
where $w_{ik}$ is the weight of $h^m_i \rightarrow y_k$, $b_k$ is the bias of $y_k$ neuron. In the following calculations, we use three hidden layers with $N$ nodes (see Table~\ref{tab:table1} for details) to represent the variational parameters in the VQE. This DNN model is accurate enough to produce the ground-state energy for our studied systems. For example, in the case of \ce{BeH2}, the DNN model of $30\times 30 \times 30$ is good enough to achieve chemical accuracy. The maximum error is even less than 0.1 millHartree when a $70\times 70 \times 70$ DNN model is used (see the Appendix A).

\begin{table*}[!htb]
    \centering
    \caption{The DNN models with three hidden layers of $N\times N \times N$ and the size of the training and testing sets used for \ce{H2}, \ce{HeH+}, \ce{H4}, \ce{LiH}, \ce{BeH2}.}
    \label{tab:table1}
    \begin{tabular}{c ccc}
    \hline\hline\noalign{\smallskip}
    Molecule & DNN Model & Training Set & Testing Set \\
    \noalign{\smallskip}\hline\noalign{\smallskip}
    \ce{H2} & $20 \times 20 \times 20$ & 36 & 16 \\
    \ce{HeH+} & $40 \times 40 \times 40$ &  65 & 30 \\
    \ce{H4} & $50 \times 50 \times 50$ &  148 & 48 \\
    \ce{LiH} & $20 \times 20 \times 20$ & 181 & 89 \\
    \ce{BeH2} & $70 \times 70 \times 70$ & 156 & 51 \\
    \noalign{\smallskip}\hline
    \end{tabular}
\end{table*}

To compute the molecular PESs, we integrate the DNN into the VQE, named DNN-VQE (see Fig.~\ref{fig:fig1}(b)). In contrast to the standard VQE, the DNN-VQE represents circuit parameters with a DNN model in order to avoid the nonlinear optimization problem in the evaluation of PESs. The DNN parameters are trained by the Adam optimizer~\cite{kingma2014adam}, and the loss function is the mean absolute percentage error (MAPE)
\begin{equation}
    \mathrm{MAPE} = \frac{1}{n} \sum^n_{t=1} \left|{\frac{y^t_{a} - y^t_{f}}{y^t_{a}}}\right| \times 100,
\end{equation}
where $n$ is the size of the training set, $y^t_{a}$ and $y^t_{f}$ are the actual output and forecast output, respectively. The training set is obtained from standard VQE calculations, in which the parameters are optimized with the Broyden-Fletcher-Goldfarb-Shannon (BFGS) algorithm. The overall flowchart for constructing the PES is shown in Fig.~\ref{fig:fig1}(c). After the DNN model is well trained, the PES is obtained from single-point energy calculations with predetermined parameterized circuits, which are expected to be very efficient on quantum devices. 

\begin{figure}[!htb]
\begin{center}
\includegraphics[width=0.7\textwidth]{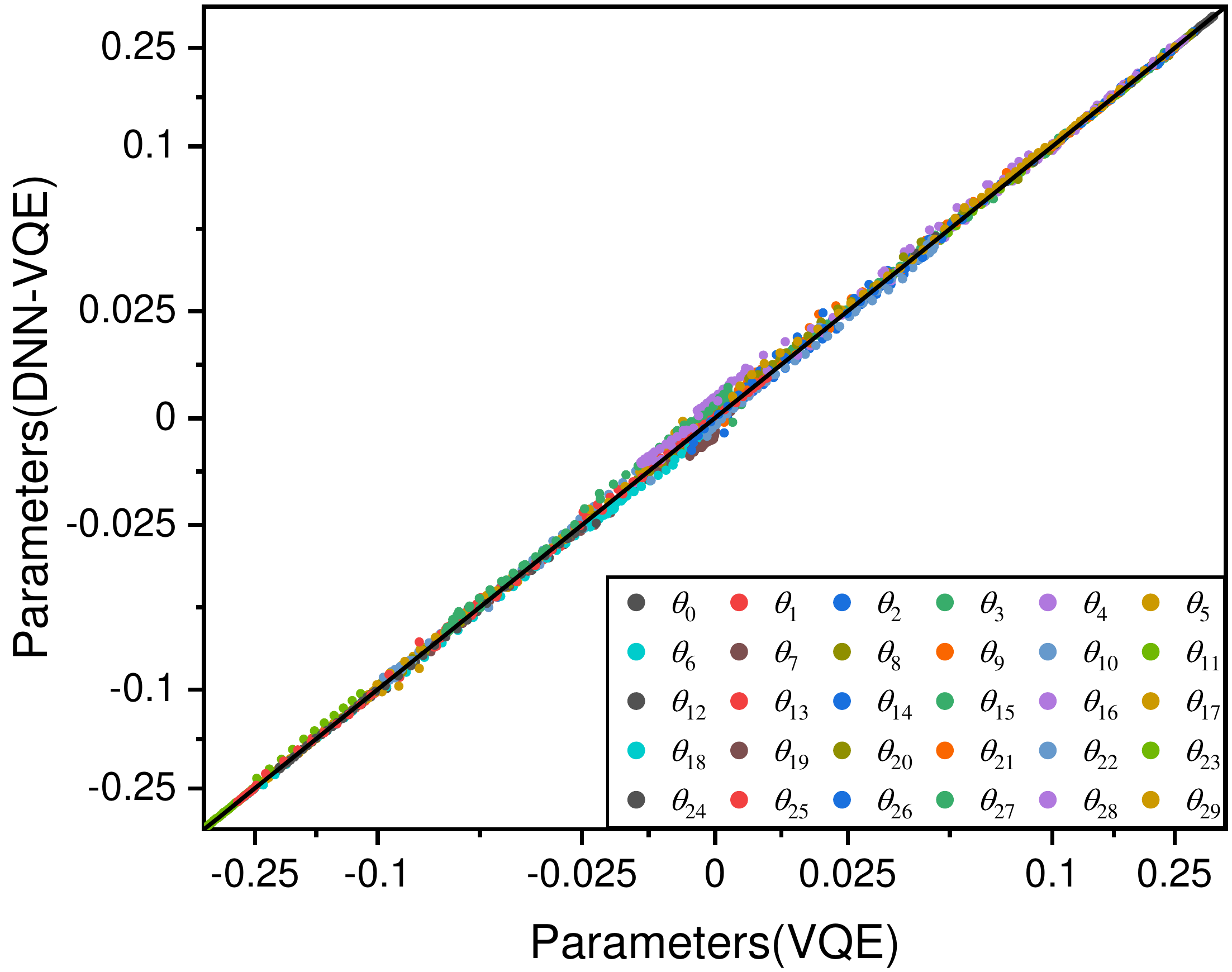}
\end{center}
\caption{Error diagram of parameters predicted by the DNN model with respect to the optimized parameters in the VQE for the \ce{H4} molecule.}
\label{fig:fig2}
\end{figure}

The hydrogen chain is a simplest realistic model to understand a variety of fundamental physical phenomena, such as an antiferromagnetic Mott phase and an insulator-to-metal transition.~\cite{MotGenMa20} We first assess the DNN-VQE method with the \ce{H4} molecule, each hydrogen atom equispaced along a line. The input parameter of the DNN model for the \ce{H4} molecule is the \ce{H-H} bond length, and outputs are 30 non-zero parameters in the VQE (Utilizing the symmetry of Hartree-Fock orbitals reduces the number of parameters of \ce{H4} from 66 to 30)~\cite{CaoHuZha21}. All non-zero parameter curves as a function of the \ce{H-H} bond length are listed in Appendix B. It is clear that variational parameters vary smoothly with respect to the \ce{H-H} bond length. The amplitudes of most parameters increase monotonically as the \ce{H-H} bond elongates. As such, a simple neural network, three hidden layers of $50 \times 50 \times 50$, is feasible for mapping the \ce{H-H} bond length to the parameter space of the VQE. Because the input vector includes only one bond length, the sizes of the training and testing sets are set to be relative small as 148 and 48, respectively. To check the reliability of the DNN model, Fig.~\ref{fig:fig2} compares parameters for the testing set predicted by the DNN model with those obtained from the standard VQE calculations. The DNN model achieves a high accuracy in predicting VQE parameters. The root-mean-square deviations of parameters for the training and testing data are $2.86 \times 10^{-3}$ and $2.74 \times 10^{-3}$, respectively, which are sufficiently small to predict the exact energy. 

\begin{figure}[!htb]
\begin{center}
\includegraphics[width=1.0\textwidth]{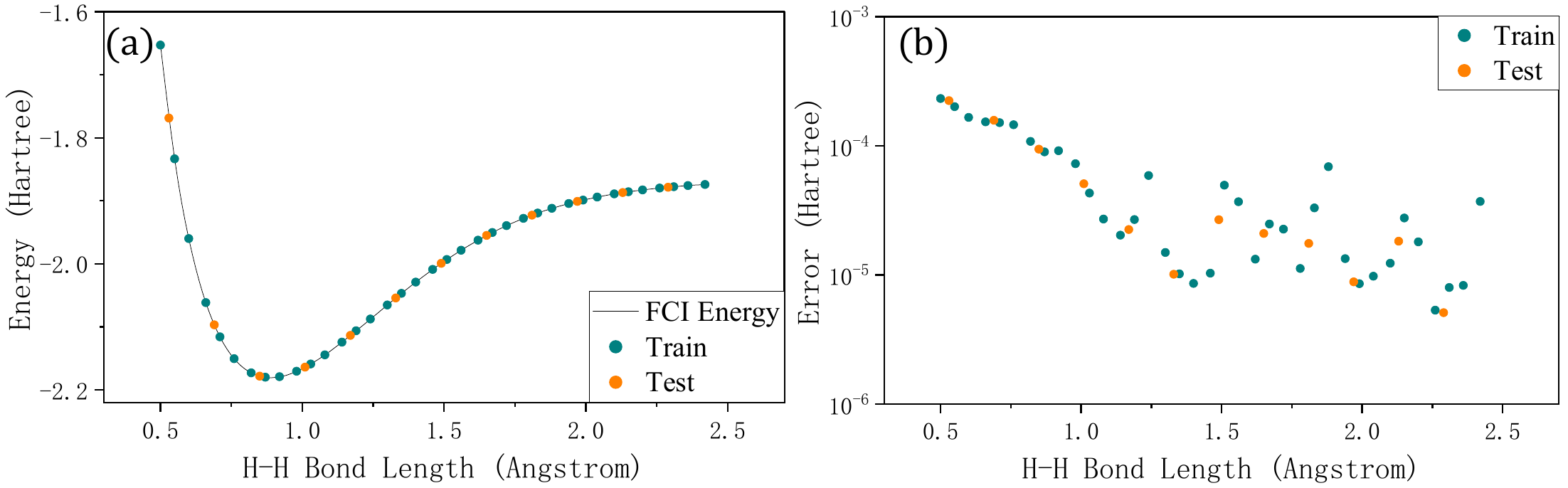}
\end{center}
\caption{(a) The ground-state energy of the \ce{H4} molecule. The black solid line is the FCI result, and the green and yellow dots represent the DNN-VQE results in the training and testing sets, respectively; (b) The errors of the DNN-VQE method with respect to the FCI result.}
\label{fig:fig3}
\end{figure}

The ground-state PES of the \ce{H4} molecule computed with the DNN-VQE method is shown in Fig.~\ref{fig:fig3}. The reference result from the full configuration interaction (FCI) method is also shown for comparison. The \ce{H-H} bond length ranges from 0.5 to 2.4 Angstrom, exhibiting a transition from weak to strong correlation effect. As discussed above, the VQE parameters generated from a DNN model agree very well with the reference values. As a consequence, the predicted ground-state PES has a indistinguishable deviation from the exact one. Most errors of the PES computed with the DNN-VQE method are on the order of magnitude of $10 ^ {-1}$ millHartree. The largest deviation of the ground-state energy between the DNN-VQE method and the FCI method (see Fig.~\ref{fig:fig3}(b)) is only 0.23 millHartree, which appears at the \ce{H-H} bond length of 0.5 Angstrom. The overlapping of the error distribution for the training and testing data reveals that the DNN model performs quite well in interpolating the VQE parameters. 

\begin{figure}[!htb]
\begin{center}
\includegraphics[width=0.7\textwidth]{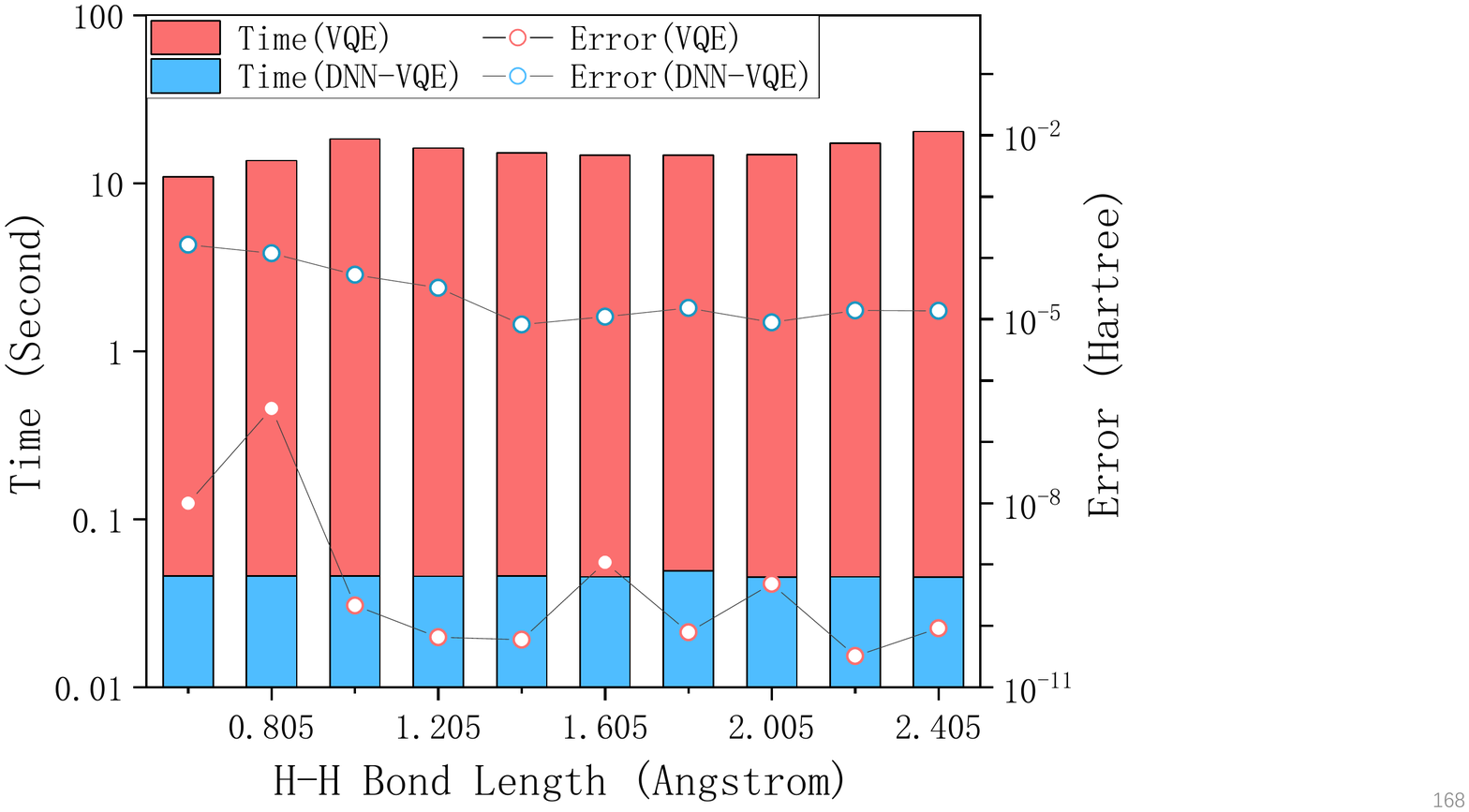}
\end{center}
\caption{The computational time of a single-point energy calculation (in second) using the standard VQE and DNN-VQE methods. Here, we exclude the computational time of Hartree-Fock calculations. The red and blue rectangles represent the time required by the standard VQE and DNN-VQE, respectively. The red and blue circles represent errors of the standard VQE and DNN-VQE, respectively.}
\label{fig:fig4}
\end{figure}

Once the DNN model is well trained, the computational time required to obtain parameters is negligible in comparison with that required to compute energies in the VQE. As such, the computational cost for obtaining the PES is significantly reduced if the energies can be efficiently computed on a quantum computer. We compare the time required to calculate the single point energy of the \ce{H4} molecule using the standard VQE and DNN-VQE method in Fig.~\ref{fig:fig4}. The average time required to compute the ground-state energies using the standard VQE algorithm is 15.64 second, while the DNN-VQE method takes only 0.05 second, 340 times speedup. The errors of the DNN-VQE method as seen from Fig.~\ref{fig:fig4} are larger than those from the standard VQE method, while they are still much less than 1 kcal/mol with parameters generated from the DNN model. 

\begin{figure}[!htb]
\begin{center}
\includegraphics[width=1.0\textwidth]{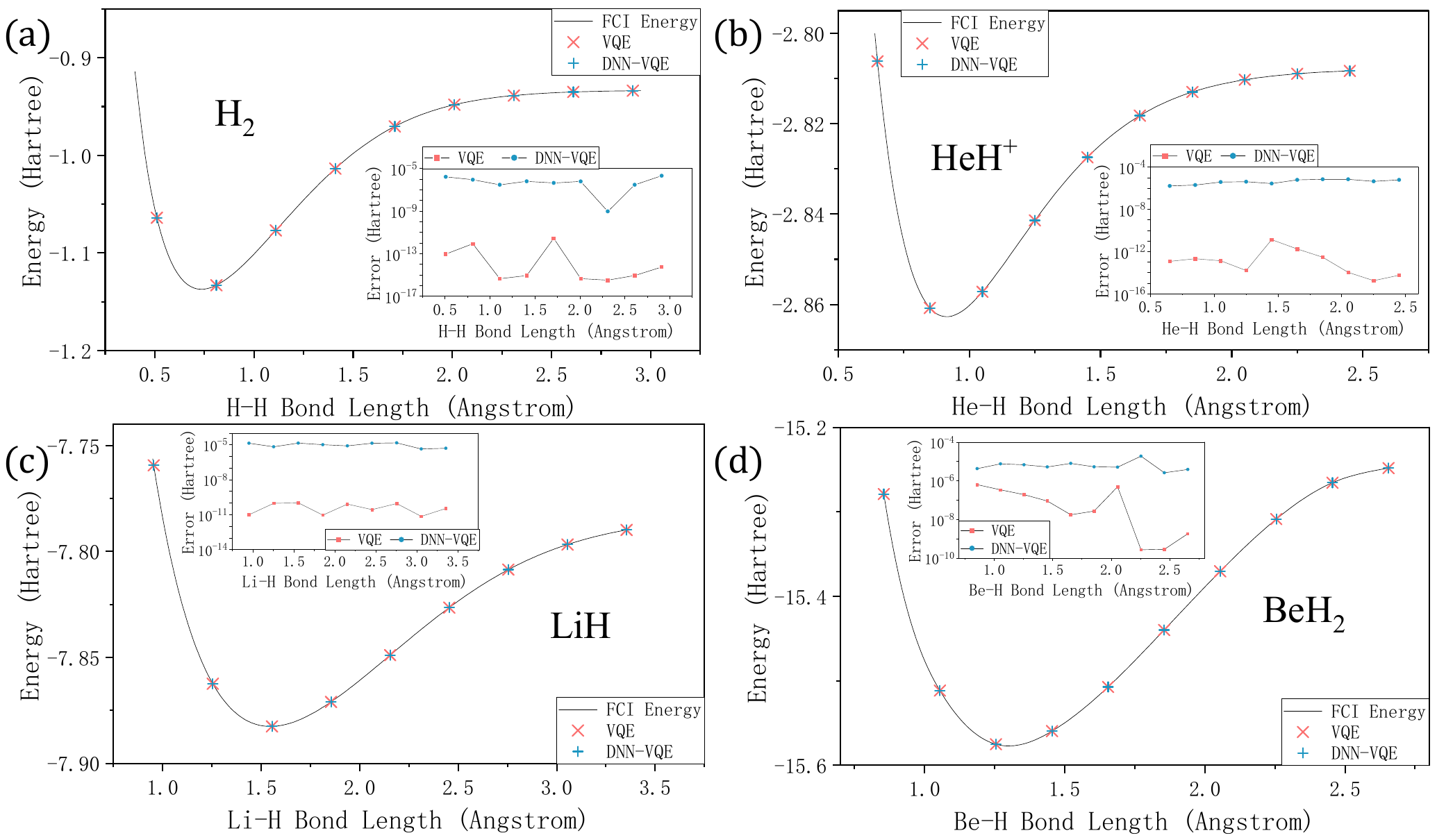}
\end{center}
\caption{Ground-state PESs of \ce{H2} (a), \ce{HeH+} (b), \ce{LiH} (c), and \ce{BeH2} (d) computed with the standard VQE, DNN-VQE and FCI methods. The black solid lines represent exact results, the red crosses represent VQE results and the blue pluses represent DNN-VQE results. The insets show errors of the DNN-VQE with respect to the exact results. The red and blue squares represent the errors of standard VQE and DNN-VQE methods, respectively.}
\label{fig:fig6}
\end{figure}

In the following, we apply the DNN-VQE method to explore the PESs of several small molecules, including \ce{H2}, \ce{HeH+}, \ce{LiH} and \ce{BeH2}. For \ce{H2} and \ce{HeH+}, four HF orbitals are included in the FCI calculations. For \ce{LiH}, the lowest HF orbital is frozen and other five HF orbitals are used in the complete active space configuration interaction (CASCI) calculations. For \ce{BeH2}, the lowest and highest HF orbitals are frozen and the active space is composed of other 5 HF orbitals. After we employ the point group symmetry to remove many excitation operators, there are 1, 4, 34, and 22 non-zero parameters left for \ce{H2}, \ce{HeH+}, \ce{LiH} and \ce{BeH2}, respectively. Fig.~\ref{fig:fig6} shows the ground-state PESs of \ce{H2}, \ce{HeH+}, \ce{LiH} and \ce{BeH2} computed with the standard VQE, DNN-VQE and exact methods. The insets show errors of the DNN-VQE with respect to the exact results. These PESs cover molecular geometries with the equilibrium bond length and bond dissociation. The maximal, minimal and average errors are listed in Table~\ref{tab:table2}. It is clear that the DNN-VQE PESs are in good agreement with the standard VQE and FCI results. The maximal errors of the DNN-VQE method for four molecules are less than $10 ^-2$ millHartree. In case of \ce{LiH} with 34 parameters, the PES computed with the DNN-VQE method is as accurate as that of \ce{H2} with 1 variational parameter. This encourages further applications of the DNN-VQE to complex systems. Overall, the DNN-VQE method is expected to provide a promising scheme for constructing accurate PESs of complex systems.

\begin{table*}[!htb]
    \centering
    \caption{Errors of the ground-state PESs computed with the DNN-VQE method (in millHartrees). $\Delta E_{max}$, $\Delta E_{min}$, and $\Delta E_{ave}$ represent the maximal, minimal and average energy errors of the DNN-VQE method, respectively.}
    \label{tab:table2}
    \begin{tabular}{c ccc}
    \hline\hline\noalign{\smallskip}
    Molecule & $\Delta E_{max}$ & $\Delta E_{min}$ & $\Delta E_{ave}$ \\
    \noalign{\smallskip}\hline\noalign{\smallskip}
    \ce{H2} & $2.13 \times 10 ^ {-3}$ & $9.60 \times 10 ^ {-7}$ & $7.78 \times 10 ^ {-4}$ \\
    \ce{HeH+} & $7.09 \times 10 ^ {-3}$ & $1.65 \times 10 ^ {-3}$ & $4.57 \times 10 ^ {-3}$ \\
    \ce{LiH} & $1.49 \times 10 ^ {-2}$ & $4.47 \times 10 ^ {-3}$ & $1.02 \times 10 ^ {-2}$ \\
    \ce{BeH2} & $1.87 \times 10 ^ {-2}$ & $2.62 \times 10 ^ {-3}$ & $6.74 \times 10 ^ {-3}$ \\
    \noalign{\smallskip}\hline
    \end{tabular}
\end{table*}

\begin{figure}[!htb]
\begin{center}
\includegraphics[width=0.7\textwidth]{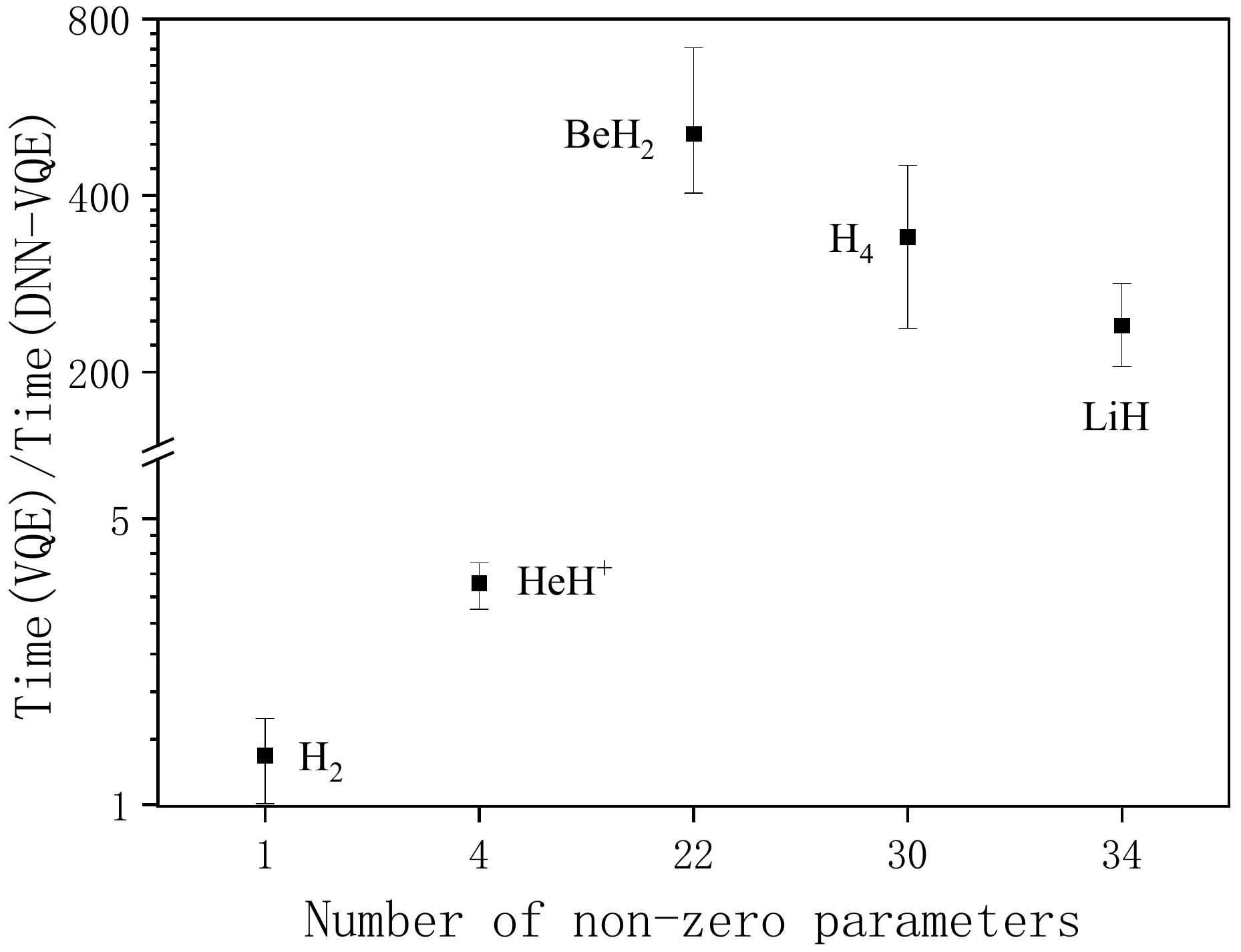}
\end{center}
\caption{Ratio of the computational time required for the standard VQE and DNN-VQE methods to compute single-point energies.}
\label{fig:fig7}
\end{figure}

For these five small molecules, we also summary the ratio of the computational time in ground-state energy calculations using the standard VQE and DNN-VQE method in Fig.~\ref{fig:fig7}. It can be seen that the DNN-VQE method significantly reduces the computational cost of a single-point energy calculation after the DNN model has been trained. The larger molecular systems are, the more computational cost we can save. For \ce{LiH} and \ce{BeH2}, the DNN-VQE method achieves hundreds of times speedup. However, in a PES calculation, we must consider the computational cost for preparing the training set, in which a large number of standard VQE calculations are required. This means that training the DNN model may be a very time-consuming step. While, if the full-dimensional PES is required, such as in dynamics simulations or predicting chemical reaction rates, the number of standard VQE calculations in the DNN-VQE method is still significantly reduced. In addition, as discussed in Ref.~\citenum{LeeHugHea19}, in order to approximate a global minimum, the optimization of parameters for the UCC ansatz requires hundreds of calculations with different initial guesses for each molecular geometry, which remarkably increases the computation cost of the VQE. The DNN-VQE method can provide a potential scheme to alleviate this problem if the DNN model can generate a good enough initial guess of parameters. More specifically, we can train a DNN model with a non-fully optimized training set, in which parameters are generated from the standard VQE calculations with a loose convergence criteria. And then the DNN model can provide an initial guess for the further VQE optimization.

In summary, we integrate the DNN and the VQE for exploring the molecular PESs. Once the DNN model is trained, the DNN-VQE method exhibits hundreds of times speedup over the standard VQE method. Therefore, the DNN-VQE method provides an appealing scheme to construct accurate PESs of complex systems if the expectation values of the Hamiltonian can be efficiently measured on a quantum computer. Note that, although the classical DNN method is employed in this work to represent variational parameters, it can be easily replaced with a quantum neutral network that has a very powerful expressivity.~\cite{WuYaoZha21} In addition, the DNN-VQE method can be straightforwardly extended to other wave function ansatzes, such as UpCCGSD and hardware-efficient ansatz.~\cite{KanMezTem17} In this work, we focus on the construction of PESs with the DNN-VQE method. While other electronic structure properties, such as charges and dipole moments, can be easily computed with the DNN-VQE method since the wave function is already known. 

\section*{Methods}

We use PySCF~\cite{sun2018pyscf} to compute one- and two-electron integrals in the second quantization Hamiltonian and use OpenFermion~\cite{mcclean2020openfermion} to carry out Jordan-Winger transformation~\cite{fradkin1989jordan}. The construction and training of neural networks in this work are performed by TensorFlow~\cite{abadi2016tensorflow}. The computational basis is the STO-3G basis set. And the wave function ansatz we used in this work is the unitary coupled cluster (UCC) ansatz~\cite{Eva11, peruzzo2014variational, SheZhaZha17, RomBabMcC18, LeeHugHea19}. For \ce{LiH}, we assume the lowest HF orbitals are always occupied. And for \ce{BeH2}, we assume the lowest HF orbitals are always occupied and the highest HF orbitals are always unoccupied. Thus, the number of qubits required in the \ce{H4}, \ce{H2}, \ce{HeH+}, \ce{LiH} and \ce{BeH2} calculations are 8, 4, 4, 10 and 10, respectively.

\section*{Acknowledgments}

This work is supported by Innovation Program for Quantum Science and Technology (2021ZD0303306), the National Natural Science Foundation of China (22073086, 21825302 and 22288201), Anhui Initiative in Quantum Information Technologies (AHY090400), and the Fundamental Research Funds for the Central Universities (WK2060000018).

\footnotesize{
\bibliography{reference}
}

\appendix

\section{The size of the DNN for \ce{BeH2} molecule}\label{sec:dnn}

\begin{figure}[H]
\begin{center}
\includegraphics[width=0.45\textwidth]{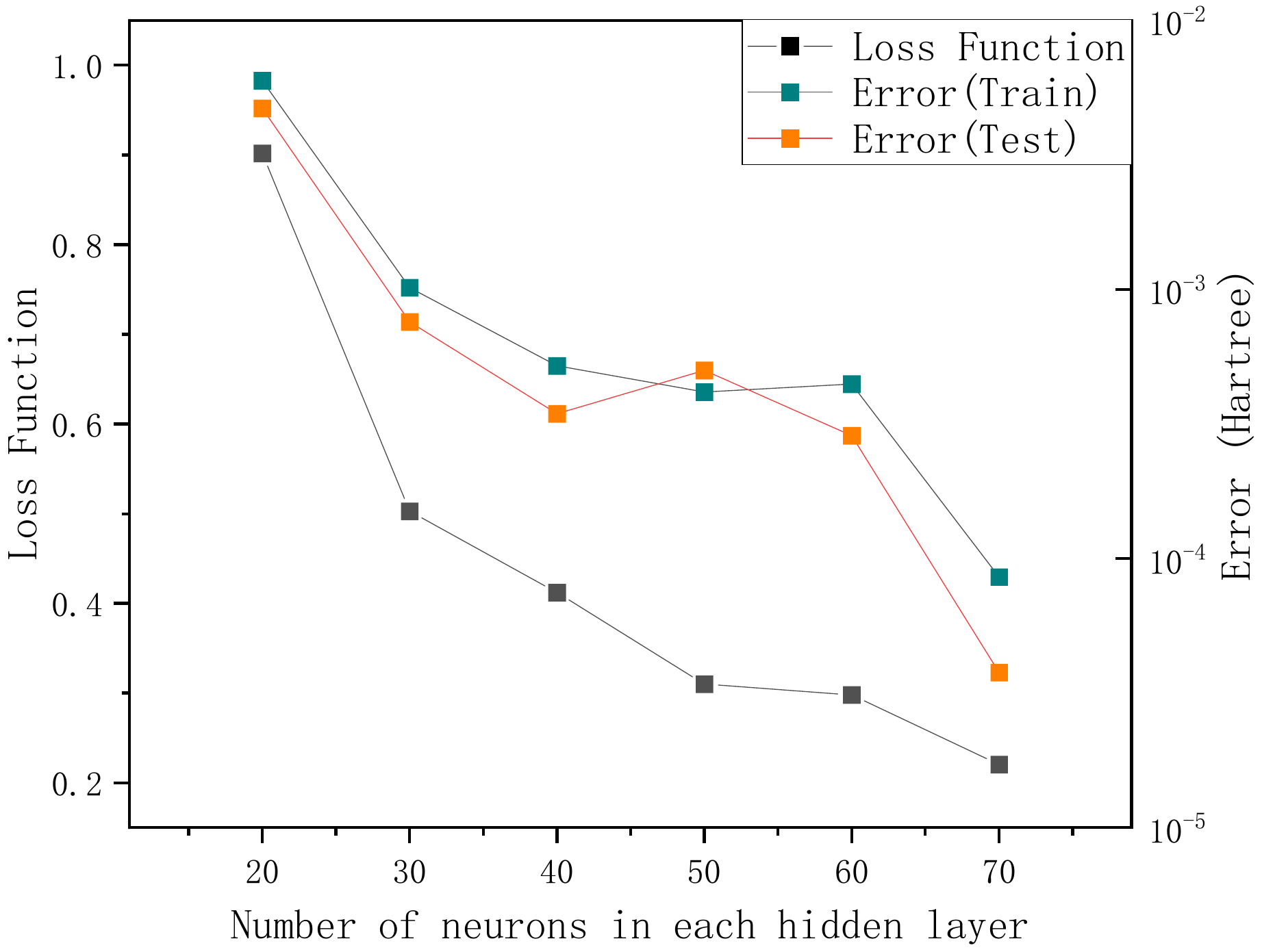}
\end{center}
\caption{The loss function and error as functions of the number of neurons in each hidden layer (N). The number of hidden layers is taken to be three, and each hidden layer has the same number of neurons. The black squares represent the loss function of the neural network. And the green and yellow squares represent the maximum training error and the maximum testing error between our results and FCI results, respectively.}
\label{fig:n}
\end{figure}

\begin{figure}[H]
\begin{center}
\includegraphics[width=0.45\textwidth]{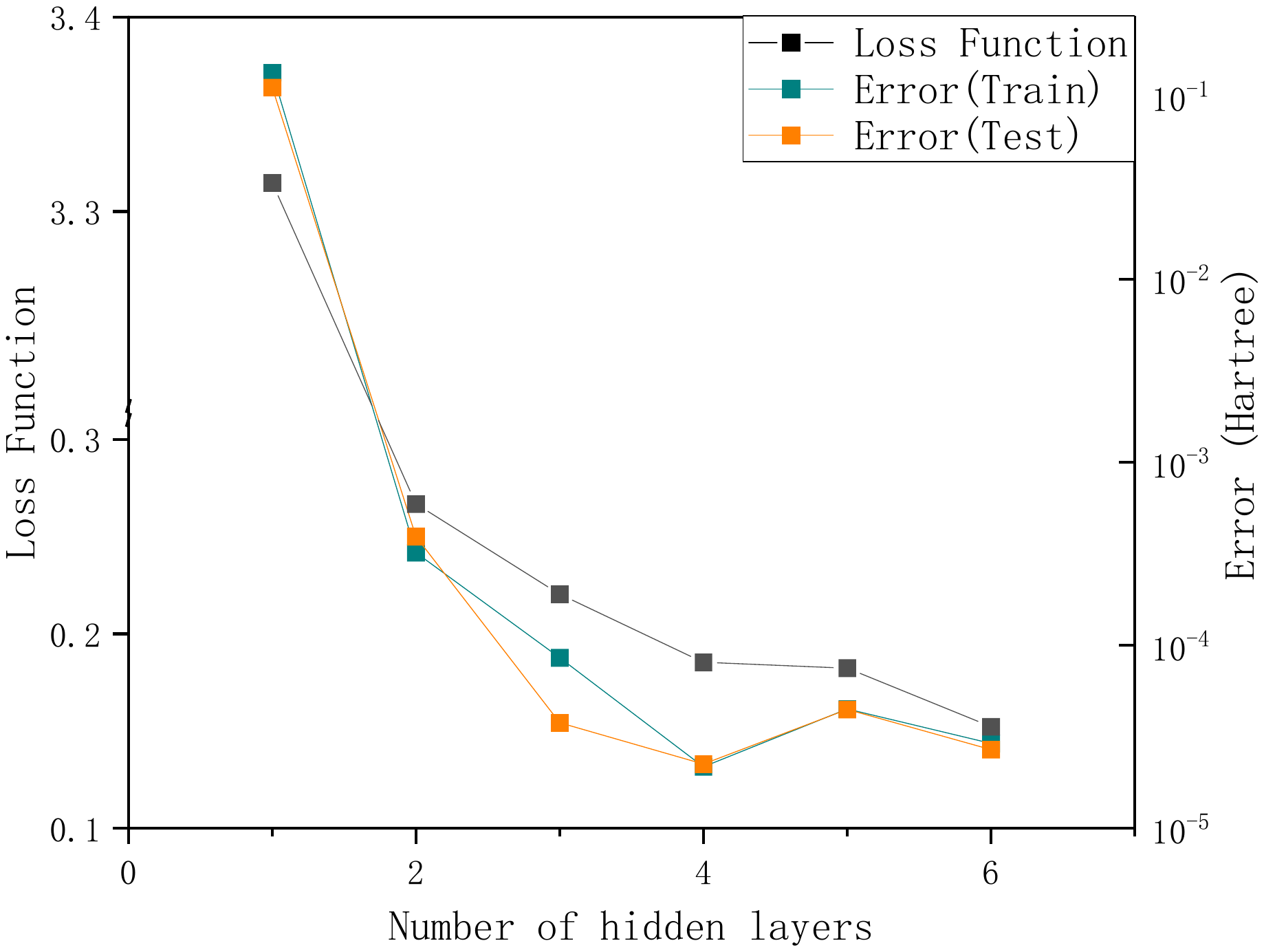}
\end{center}
\caption{The loss function and error as functions of the number of hidden layers (L). The number of neurons in each hidden layer is taken to be 70. The black squares represent the loss function of the neural network. And the green and yellow squares represent the maximum training error and the maximum testing error between our results and FCI results, respectively.}
\label{fig:l}
\end{figure}

In order to determine the size of the deep neural network (DNN), Fig.~\ref{fig:n} and Fig.~\ref{fig:l} depict the numerical results for \ce{BeH2}, with different numbers of neurons in each hidden layer (N) and hidden layers. Since the values of variational parameters for \ce{BeH2} are too small, we use $\frac{1}{parameter + 1}$ as the output of the neural network to accelerate the convergence of the loss function. The black squares in Fig.~\ref{fig:n} show that the loss function of the DNN decreases as N increases. When N increases to 30, the results can achieve chemical accuracy. And when N increases to 70, the maximum error for \ce{BeH2} is in the order of $10^{-2}$ millHartree. For the sake of high accuracy, we take N to be 70 for \ce{BeH2} molecule.

\begin{figure}[H]
\begin{center}
\includegraphics[width=0.45\textwidth]{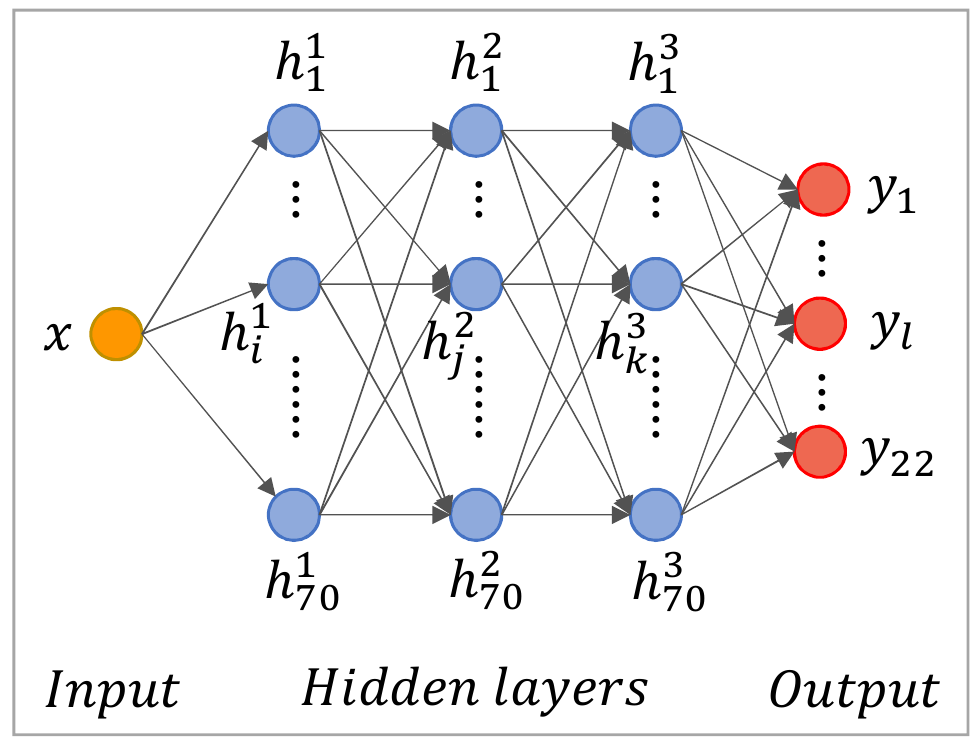}
\end{center}
\caption{Schematic model of the deep neural network used for \ce{BeH2} in this work. It is a single input and multiple output neural network with three hidden layers. And there are 70 neurons in a hidden layer. Yellow circle, blue circles and red circles represent neurons in the input, hidden and output layers, respectively.}
\label{fig:dnn-beh2}
\end{figure}

Then, we test the effect of the number of hidden layers (L) in Fig.~\ref{fig:l}. The black squares show that the loss function of the DNN also decreases as L increases. When L is equal to 2, the results can achieve chemical accuracy. And when L increases to 3, the order of magnitude of the maximum error for \ce{BeH2} is $10^{-2}$ millHartree. Since the accuracy does not improve significantly as L continues to increase, we use a three hidden layers DNN with 70 neurons in each hidden layer for \ce{BeH2}, which is illustrated in Fig.~\ref{fig:dnn-beh2}. 

\section{Variational parameters of the UCCGSD ansatz}\label{sec:parameter}

\begin{figure}[H]
\begin{center}
\includegraphics[width=0.45\textwidth]{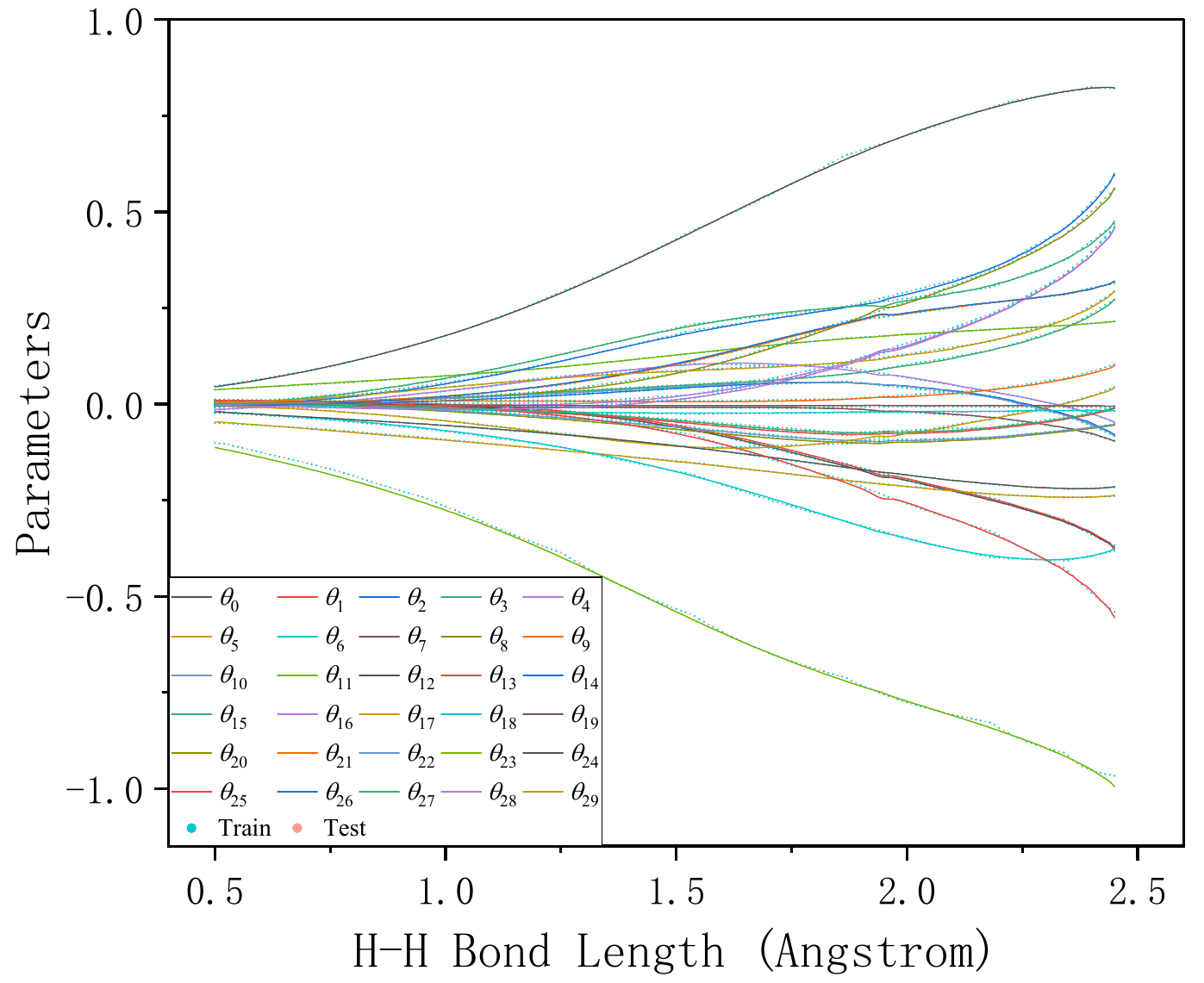}
\end{center}
\caption{Variational parameters of the UCCGSD ansatz as functions of the \ce{H-H} bond length for the \ce{H4} molecule. Utilizing the symmetry of Hartree-Fock molecular orbitals, we reduce the number of variational parameters of \ce{H4} to 30. Color curves represent parameters $\vec{\theta}$ obtained from the classical optimizer. The blue and pink dots represent the predicted results of the training set and testing set by the DNN, respectively.}
\label{fig:parameter-h4}
\end{figure}

\begin{figure}[H]
\begin{center}
\subfigure[]{\includegraphics[width=0.45\textwidth]{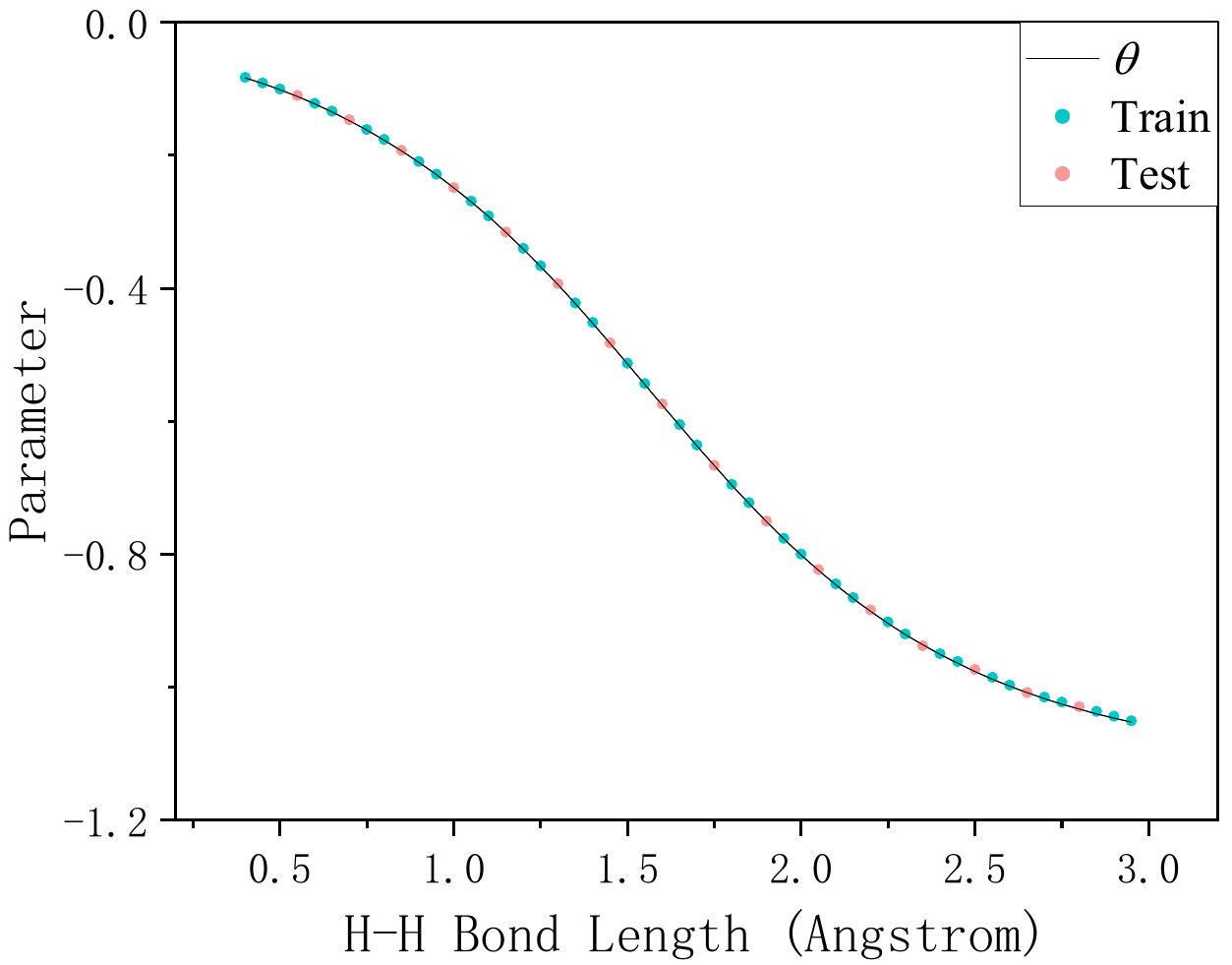}}
\subfigure[]{\includegraphics[width=0.45\textwidth]{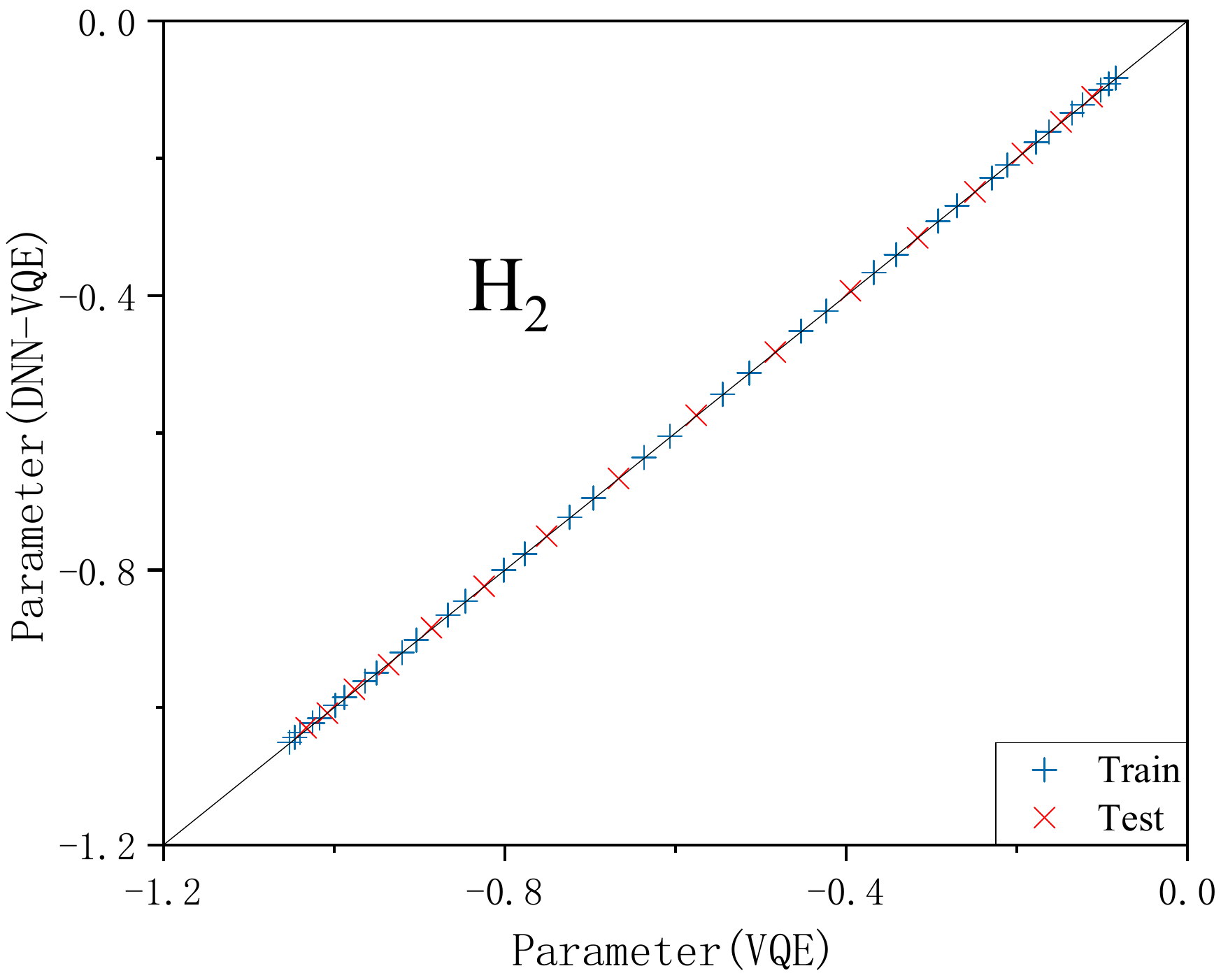}}
\end{center}
\caption{
(a) Variational parameter of the UCCGSD ansatz as a function of the \ce{H-H} bond length for the \ce{H2} molecule. \ce{H2} has one non-zero variational parameter. The black line represents the parameter obtained from the classical optimizer. The blue and pink dots represent the predicted results of the training set and testing set by the DNN, respectively;
(b) Error diagram of variational parameter predicted by the DNN model with respect to the optimized parameter in the VQE for the \ce{H2} molecule. The abscissa is the variational parameter obtained in the standard VQE, and the ordinate is the variational parameter predicted by the DNN model.The pluses and the crosses represent the results from the training set and testing set, respectively.}
\label{fig:parameter-h2}
\end{figure}

\begin{figure}[H]
\begin{center}
\subfigure[]{\includegraphics[width=0.45\textwidth]{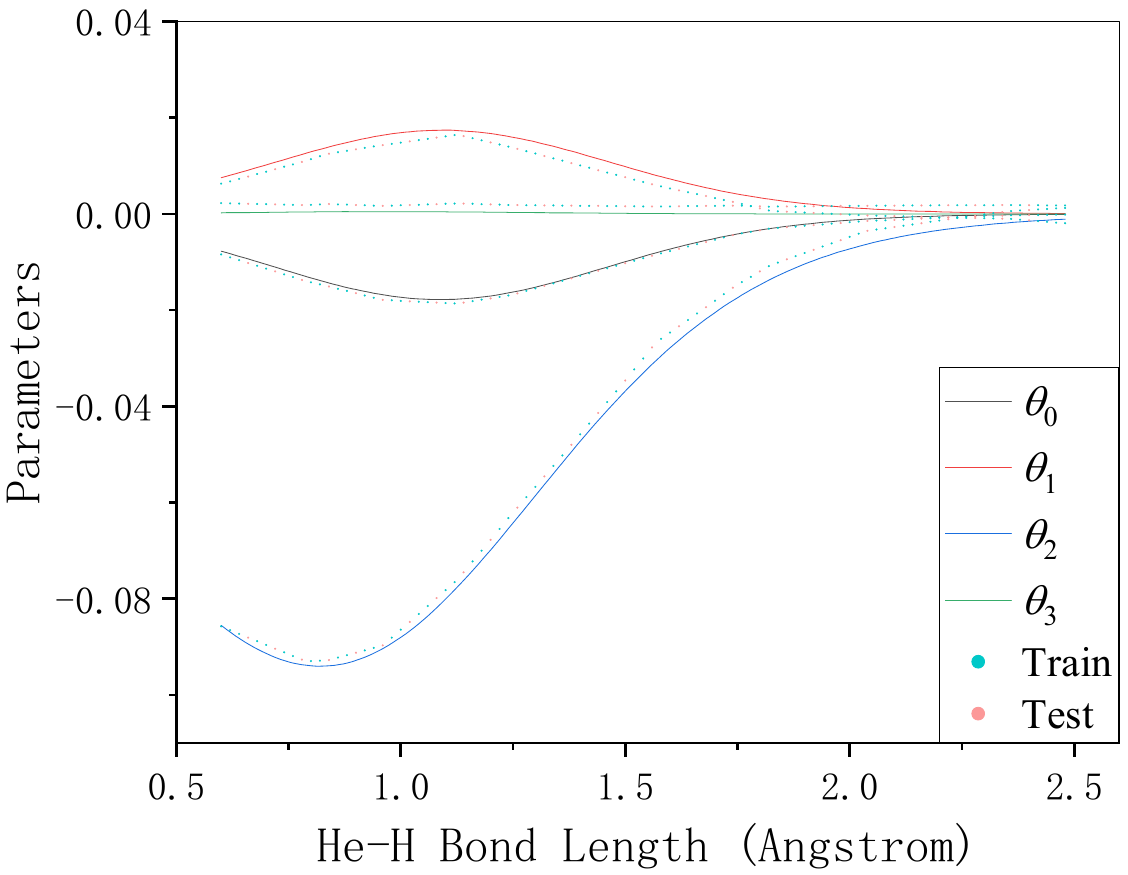}}
\subfigure[]{\includegraphics[width=0.45\textwidth]{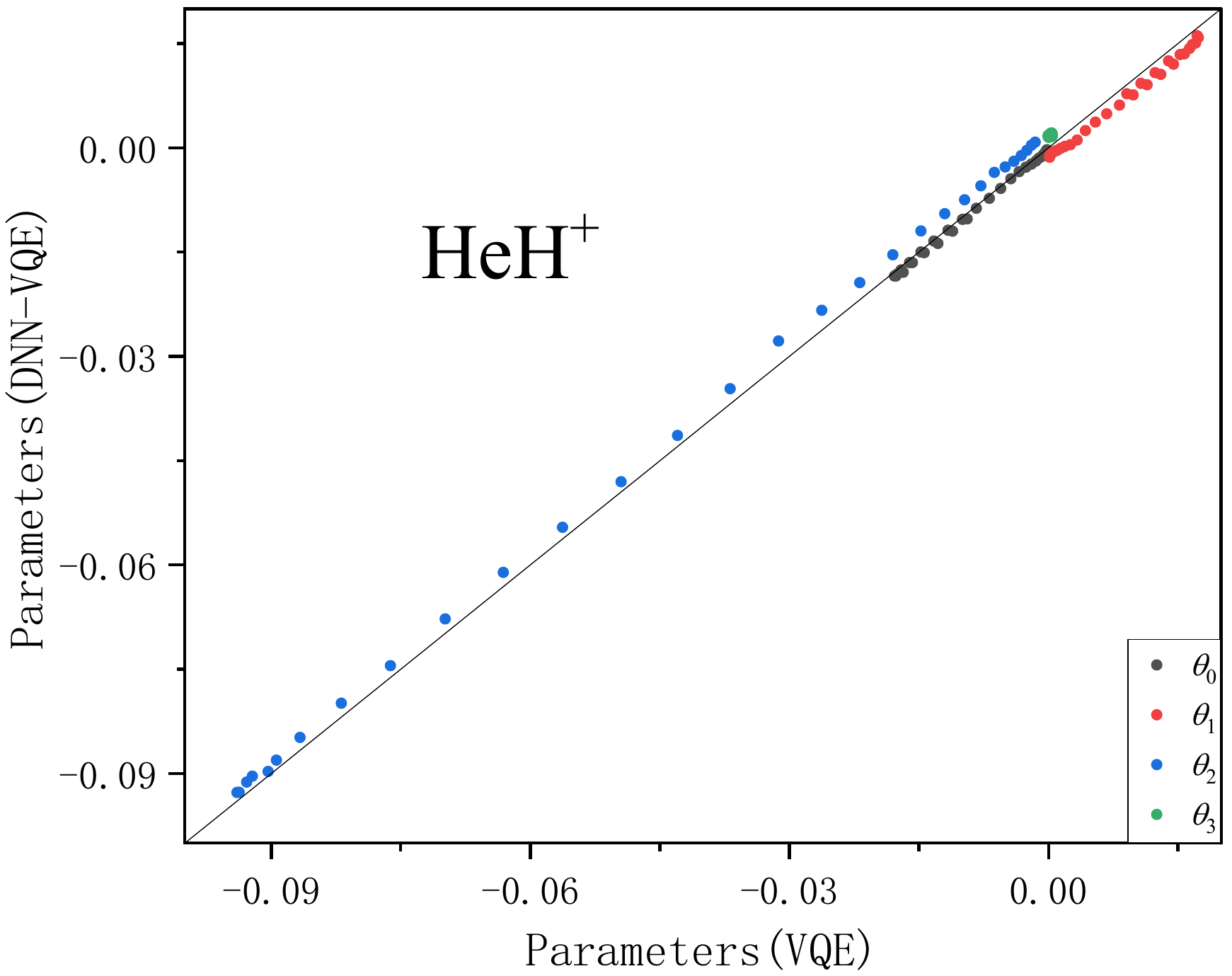}}
\end{center}
\caption{
(a) Variational parameters of the UCCGSD ansatz as functions of the \ce{He-H} bond length for the \ce{HeH+} ion. \ce{HeH+} has 4 non-zero variational parameters. Color curves represent the parameters $\vec{\theta}$ obtained from the classical optimizer. The blue and pink dots represent the predicted results of the training set and testing set by the DNN, respectively;
(b) Error diagram of variational parameters predicted by the DNN model with respect to the optimized parameters in the VQE for the \ce{HeH+} molecule (testing set). The abscissa is the variational parameters obtained in the standard VQE, and the ordinate is the variational parameters predicted by the DNN model.}
\label{fig:parameter-heh+}
\end{figure}

\begin{figure}[H]
\begin{center}
\subfigure[]{\includegraphics[width=0.45\textwidth]{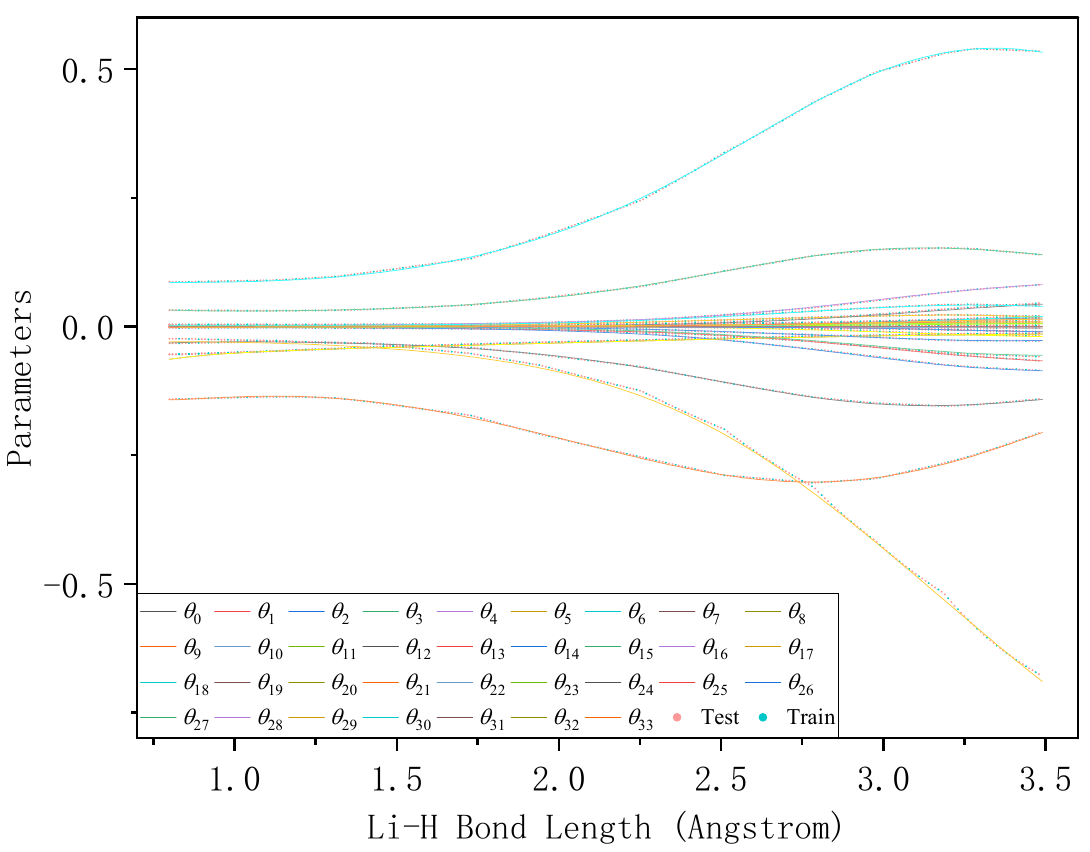}}
\subfigure[]{\includegraphics[width=0.45\textwidth]{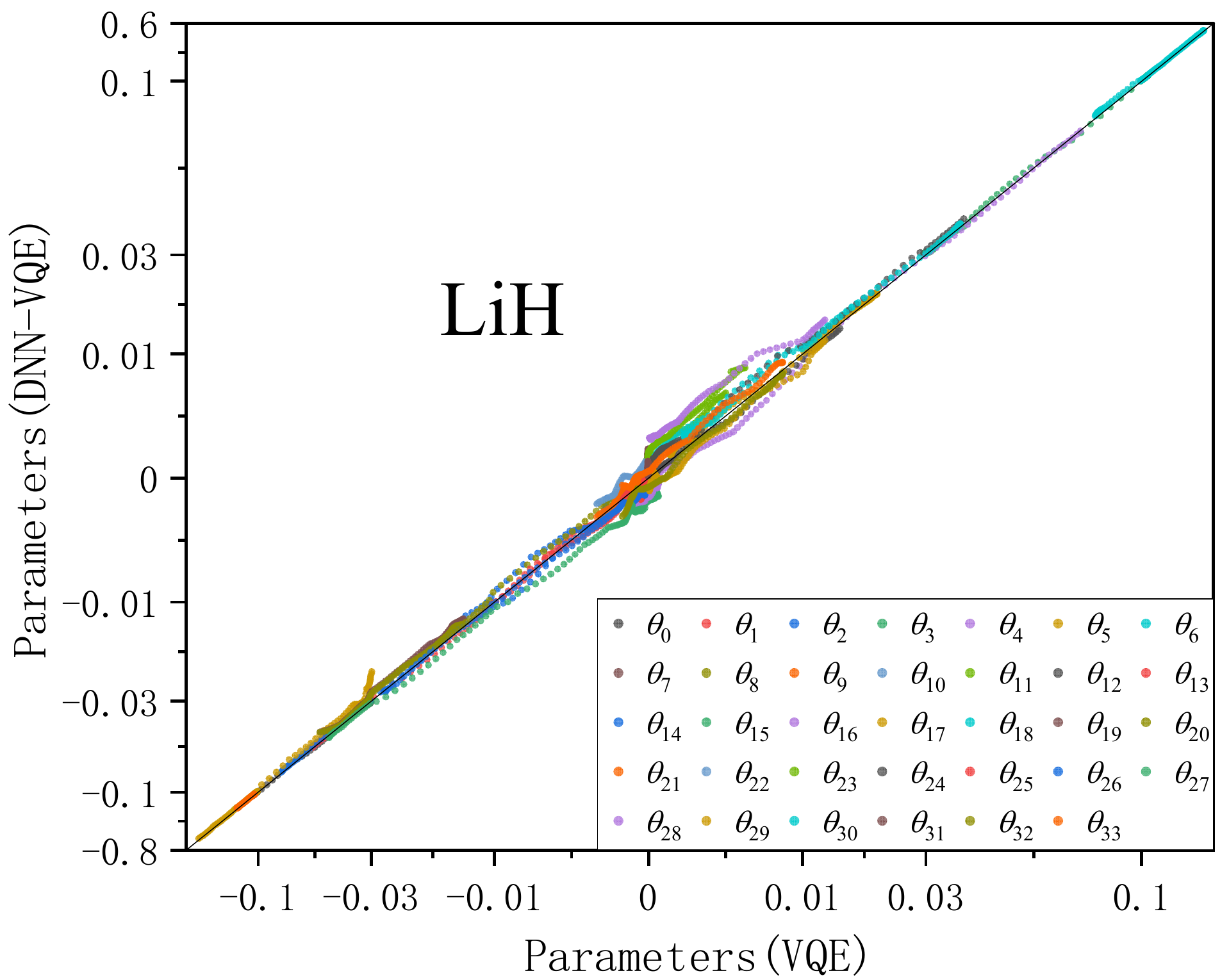}}
\end{center}
\caption{
(a) Variational parameters of the UCCGSD ansatz as functions of the \ce{Li-H} bond length for the \ce{LiH} molecule. \ce{LiH} has 34 non-zero variational parameters. Color curves represent the parameters $\vec{\theta}$ obtained from the classical optimizer. The blue and pink dots represent the predicted results of the training set and testing set by the DNN, respectively;
(b) Error diagram of variational parameters predicted by the DNN model with respect to the optimized parameters in the VQE for the \ce{LiH} molecule (testing set). The abscissa is the variational parameters obtained in the standard VQE, and the ordinate is the variational parameters predicted by the DNN model.}
\label{fig:parameter-lih}
\end{figure}

\begin{figure}[H]
\begin{center}
\subfigure[]{\includegraphics[width=0.45\textwidth]{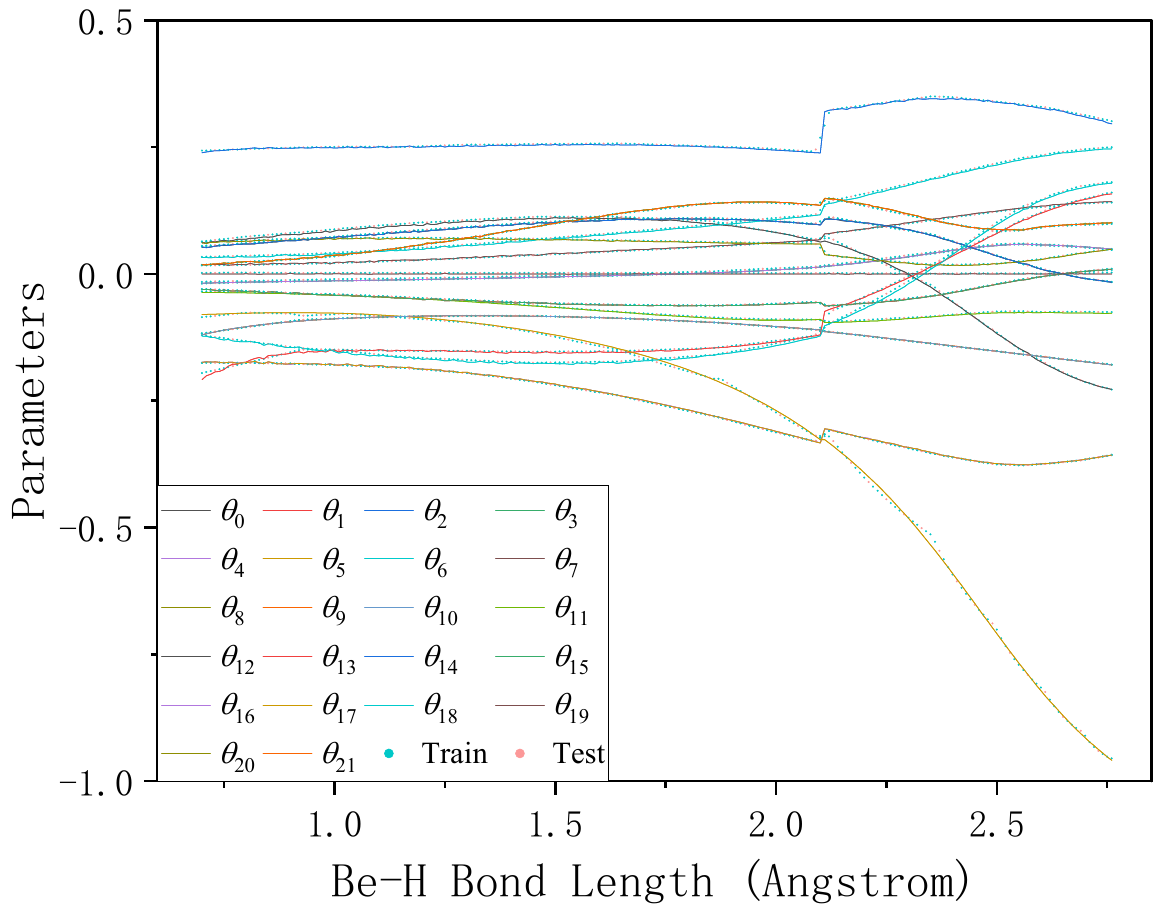}}
\subfigure[]{\includegraphics[width=0.45\textwidth]{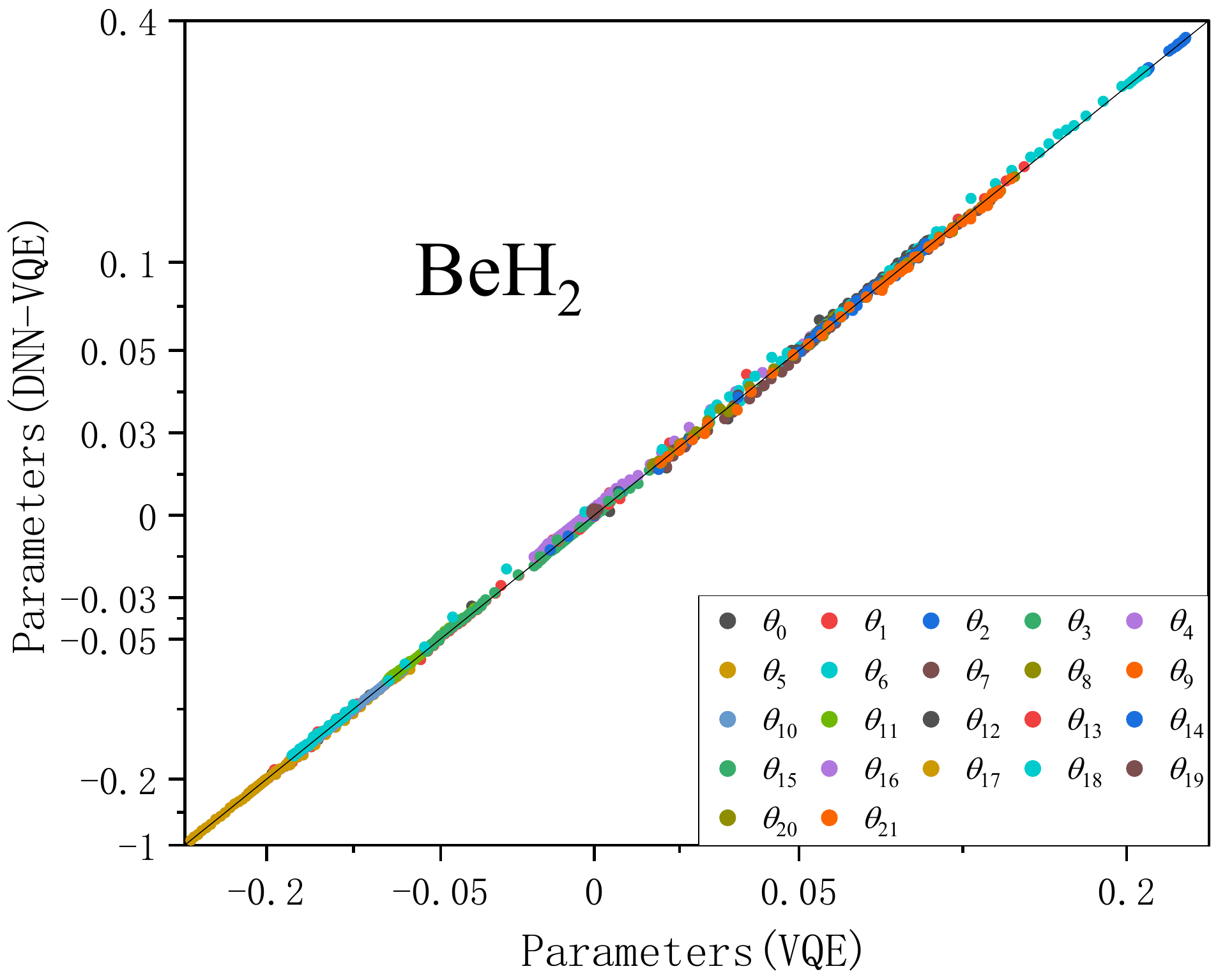}}
\end{center}
\caption{
(a) Variational parameters of the UCCGSD ansatz as functions of the \ce{Be-H} bond length for the \ce{BeH2} molecule. \ce{BeH2} has 22 non-zero variational parameters. Color curves represent the parameters $\vec{\theta}$ obtained from the classical optimizer. The blue and pink dots represent the predicted results of the training set and testing set by the DNN, respectively;
(b) Error diagram of variational parameters predicted by the DNN model with respect to the optimized parameters in the VQE for the \ce{BeH2} molecule (testing set). The abscissa is the variational parameters obtained in the standard VQE, and the ordinate is the variational parameters predicted by the DNN model.}
\label{fig:parameter-beh2}
\end{figure}

\end{document}